# An In-depth Analysis of a Cyber Attack: Case Study and Security Insights

**Puya Pakshad**
**Illinois Institute of Technology, College of Computing,**
**10 West 35th Street, Chicago, IL 60616, USA**

**Abstract**:
Nation-sponsored cyberattacks pose a significant threat to national security by targeting critical infrastructure and disrupting essential services. One of the most impactful cyber threats affecting South Korea's banking sector and infrastructure was the DarkSeoul cyberattack, which occurred several years ago. Believed to have been orchestrated by North Korea state-sponsored hackers, the attack employed spear phishing, DNS poisoning, and malware to compromise systems, causing widespread disruption. In this paper, we conduct an in-depth analysis of the DarkSeoul attack, examining the techniques used and providing insights and defense recommendations for the global cybersecurity community. The motivations behind the attack are explored, along with an assessment of South Korea's response and the broader implications for cybersecurity policy. Our analysis highlights the vulnerabilities exploited and underscores the need for more proactive defenses against state-sponsored cyber threats. This paper, therefore, emphasizes the critical need for stronger national cybersecurity defenses in the face of such threats.

## Introduction

Cyber-attacks are increasingly reported worldwide, causing significant damage to software, networks, data, and infrastructure. Studying and analyzing these attacks enables us to better understand the technical damages and gain valuable experience, allowing for more effective strategies in defense and countermeasures. One particularly dangerous cyberattack took place in South Korea during the final days of June 2013, coinciding with the anniversary of the start of the Korean War in 1950. The hackers appeared to mark occasion by disrupting websites associated with the South Korean president's office and several local newspapers. Although government officials have not officially attributed the attacks, other sources have linked at least one of them to a group known as DarkSeoul, which has been targeting South Korea for four years (Prince, 2013; Trim & Lee, 2010). This group is believed to be connected to the cyberattacks in March that resulted in the wiping of numerous hard drives at South Korean banks and television stations, as well as the more recent attacks on financial companies (Trim & Lee, 2010). The DarkSeoul cyberattack, which targeted major South Korean banks and broadcasting companies, is regarded as one of the most impactful cyber threats in the country's history. Allegedly orchestrated by North Korean state-sponsored hackers, this attack employed a range of malicious techniques, including spear-phishing, DNS poisoning, and destructive malware, to paralyze critical infrastructure. Over 48,000 computers were rendered inoperable, resulting in significant disruption of essential services for at least one day (Martin, 2016; Marpaung & Lee, 2013; Coleman, 2010). The attack's relatively low technical sophistication—compared to threats such as Stuxnet or 10 Days of Rain—was overshadowed by its high impact on South Korea's financial and media sectors. Despite using recycled malware from previous operations like Operation Troy, the attackers managed to penetrate multiple networks, pivot to critical systems, and erase data from both Windows and Unix-like operating systems. This paper aims to provide an in-depth analysis of the DarkSeoul attack, focusing on the techniques used, the impact on South Korea's infrastructure, and the broader cybersecurity lessons learned from the incident.

In general, the key focus areas of this paper are as follows:

- Provide an in-depth analysis of the DarkSeoul cyberattack scenario, including a comparison with other similar cyber warfare attacks.

- Analyze the South Korean government's response to the attack and its efforts to strengthen national cybersecurity defenses.

- Conduct a thorough and comprehensive exploration of the techniques employed in the DarkSeoul attack.

- Propose our recommendations for proactive strategies to prevent future state-sponsored cyberattacks.

The rest of the paper is organized as follows. Section 2 provides a background review of the DarkSeoul cyberattack, including its technical details and motivations. Section 3 presents a comparative analysis between the DarkSeoul attack and other similar nation-sponsored cyberattacks. Section 4 discusses the techniques used in the DarkSeoul attack and offers insights for future defense strategies. Finally, Section 5 provides conclusions and recommendations for proactive measures against state-sponsored cyberattacks.

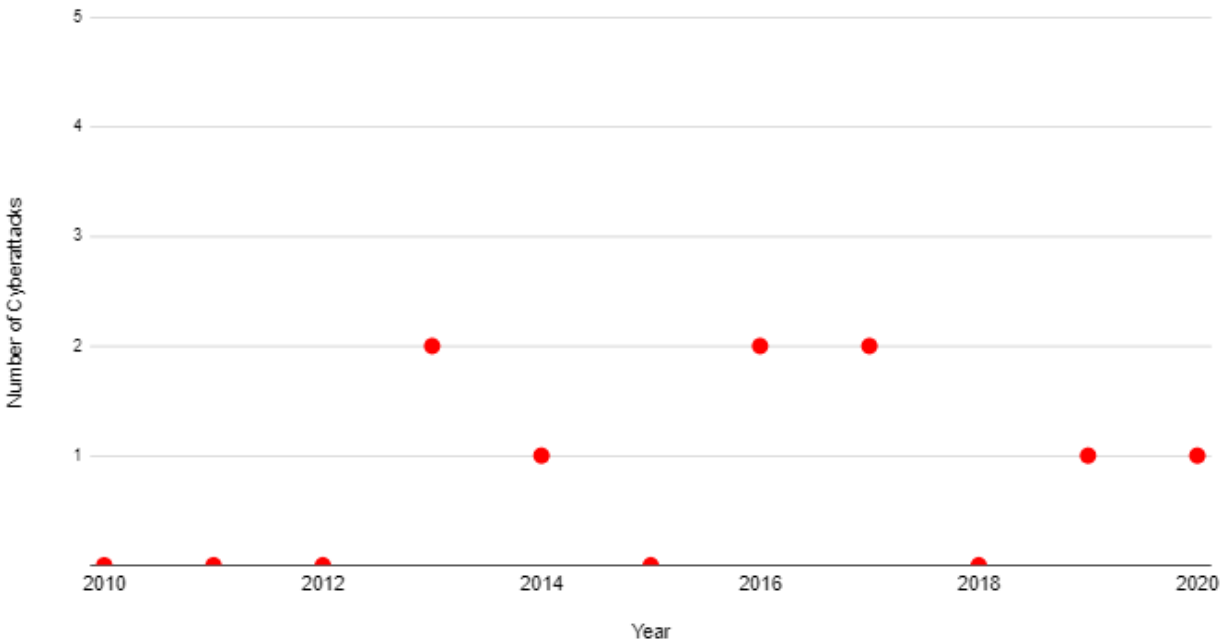

Figure. 1: Number of Suspected Cyber-attacks Linked to North Korea from 2010-2020

**Background**

Before delving into the analysis of the DarkSeoul attack, this section will provide an overview of malware within the cybersecurity domain. Malware, or malicious software, is a broad term that refers to any software intentionally designed to cause harm to a computer system, network, or device. It can take many forms, including viruses, worms, Trojan horses, ransomware, and spyware. The primary goal of malware is often to steal sensitive information, damage systems, or disrupt services, typically for financial gain, political motives, or personal reasons. In recent years, the complexity and frequency of malware attacks have increased, making them a significant threat to businesses, governments, and individuals worldwide. Three common motivations drive cyber attacks: political, criminal, and personal. Politically motivated attacks often aim to disrupt or destabilize government entities or to protest certain policies (Trim & Lee, 2010; Nah, 2023; Shin, Lee, & Kim, 2018). Criminal attacks focus on financial gain, such as stealing sensitive data or deploying ransomware. Personal attacks, on the other hand, can be driven by disgruntled employees or individuals seeking revenge. Regardless of the motive, attackers frequently use malware to exploit vulnerabilities and cause system-wide damage.

Understanding the nature and goals of malware is crucial for recognizing its role in various types of cyber attacks. In this context, examining the methods employed in the DarkSeoul attack offers valuable insights into the evolving landscape of cyber warfare (North Korea's Cyber Operations, 2016; Park & Kim, 2024). This case illustrates how state-sponsored groups increasingly leverage relatively unsophisticated yet highly destructive techniques to achieve political and economic disruption. As such, it underscores the growing need for robust defensive strategies to safeguard critical infrastructure from similar threats. By analyzing such attacks, we can better appreciate the importance of advanced security measures in defending against emerging cyber risks (Desimone & Horton, 2017; Coleman, 2010).

**Impact of DarkSeoul**

The DarkSeoul cyberattack was a highly destructive event that revealed critical weaknesses in the nation's digital infrastructure. By targeting essential services such as banking and media, the attack demonstrated the far-reaching consequences of cyber warfare, where disruptions in key sectors can ripple across the entire economy. This incident showcased the vulnerability of both public and private institutions, which rely heavily on interconnected systems that, without strong security measures, can become single points of failure. Compounding these vulnerabilities is the reality that computer systems across various layers—software applications, operating systems, networks, and infrastructures—remain susceptible to security breaches(Pakshad et al., 2023; Sultana et al., 2023; Papain, 2011; Rossum, 2017). This is largely due to insufficient integration of comprehensive security policies during the development phase. In many cases, software vendors do not provide adequate documentation for assessing non-functional requirements, and security considerations are often limited to the basic principles of confidentiality, integrity, and availability (Pakshad et al., 2023; Sultana et al., 2023; Raska, 2022). These shortcomings create exploitable weaknesses, allowing attackers, like those responsible for DarkSeoul, to penetrate systems with relative ease. As this attack demonstrated, the failure to address such vulnerabilities during development can lead to severe consequences when systems are targeted. The DarkSeoul attack had both technical and psychological impacts. Technically, it caused widespread failures across crucial infrastructure systems, while psychologically, it undermined public trust in the government's ability to protect essential services. In the following sections, we will explore how the attack immediately affected key infrastructure and the broader implications it holds for cybersecurity strategy.

A. Immediate Impact on Critical Infrastructure

The DarkSeoul cyberattack had a significant and immediate impact on South Korea's critical infrastructure, particularly in the banking and media sectors. Approximately 48,000 systems were affected across key institutions, including Shinhan Bank, Nonghyup Bank, and major media outlets such as YTN, KBS, and MBC. Services were severely disrupted for over 24 hours, with widespread outages affecting ATMs, mobile banking, and online financial services, leaving millions of South Koreans without access to essential resources. The attack, which rendered systems inoperable, created substantial public panic and resulted in considerable economic losses. This attack was both technical and psychological, aiming to undermine public trust in the country's digital infrastructure.

B. Broader Cybersecurity Implications

Figure 1 illustrates the number of cyberattacks attributed to North Korea as the perpetrator. It is important to note, however, that while some of these cyberattacks are officially linked to North Korea, others remain suspected rather than confirmed. As of 2024, the United Nations has identified North Korea as the suspected actor behind 58 attacks that are believed to have supported its nuclear weapons program. One notable cyber operation, attributed to North Korea state-sponsored actors, was designed to inflict both material and emotional damage. Although initially perceived as a low-sophistication attack due to its reliance on destructive malware, the scale of its impact revealed significant vulnerabilities within South Korea's cybersecurity defenses. By disrupting key financial services and media broadcasts, the attack highlighted the fragility of national infrastructure and demonstrated North Korea's growing cyber capabilities. This operation served a strategic purpose, aiming to erode confidence in the government's ability to protect critical infrastructure and amplify fear among the population (Martin, 2016; Jun et al., 2014; Kim, 2022).

**Authorities' Challenges**

Attributing the responsibility for the DarkSeoul cyberattack posed considerable challenges for South Korean authorities. While North Korea was the prime suspect, the country repeatedly denied involvement, and the attackers employed tactics that complicated the investigation. They reused malware from previous campaigns, including Operation Troy (2009–2013), which had targeted South Korea in earlier cyber threats. This reuse of malware, alongside the use of anonymizing techniques such as proxy servers and TOR exitnodes, obscured the origin of the attack, blurring lines of accountability (Martin, 2016; Mavroeidis et al., 2021; Iftikhar, 2024). Despite these obstacles, South Korean investigators ultimately attributed the attack to North Korea. The malware used in the DarkSeoul attack bore strong similarities to tools deployed in earlier cyber campaigns, and network logs revealed that some of the infrastructure used in the attack was linked to North Korean IP addresses. However, the attackers' use of proxy servers and anonymized networks made it difficult to conclusively trace the attack exclusively to North Korea (Martin, 2016; Mavroeidis et al., 2021; Iftikhar, 2024). A crucial part of the attack involved exploiting patch management systems, allowing the attackers to distribute malware across entire networks with minimal detection. This strategy of using legitimate infrastructure increased the complexity of both detecting and mitigating the threat. Investigators also noted that maintaining such a wide-reaching network of compromised infrastructure required significant time and resources, further implicating state-sponsored actors (Martin, 2016; Marpaung & Lee, 2013; Iftikhar, 2024;

Siers, 2014). According to the spokesman for the Korea Internet and Security Agency (KISA), 22 of the IP addresses used in the DarkSeoul attack had been involved in previous attacks attributed to North Korean hackers since 2009 (Martin, 2016). The use of anonymizing tools and sophisticated malware distribution tactics demonstrated the attackers' operational sophistication. These challenges underscored the need for South Korea to strengthen its cyber-security defenses. In response to the growing threat of state sponsored cyber operations, South Korea established the Cyber Command, marking a significant step in enhancing its national cyber defense posture (Martin, 2016; Park, Rowe, & Cisneros, 2016; Chah, 2014).

A. International Response

The DarkSeoul cyberattack sent shockwaves through both national and international communities, emphasizing the growing need for cohesive and robust cybersecurity measures. While the attack was focused on a single nation, its broader implications highlighted the vulnerability of global interconnected systems and the potential for cyberattacks to destabilize critical infrastructures. This event served as a wake-up call for both national governments and the international community, urging them to rethink and strengthen their cybersecurity frameworks.

At the national level, the attack forced governments to consider the importance of both defensive and offensive cybersecurity capabilities. Defensive measures ensure that critical infrastructures such as financial systems, media networks, and public services are adequately protected from disruptive attacks. However, the attack also demonstrated the necessity of developing offensive capabilities to counter and potentially deter future state-sponsored cyberattacks (Baezner, 2018; Sultana, Boyd, & Williams, 2023; Kim, Alfouzan, & Kim, 2021; Nah, 2023). On the international stage, DarkSeoul highlighted the increasing prevalence of cyber warfare and the challenges associated with attributing attacks to specific state actors. The complexity of attribution made it difficult to respond decisively, but the attack ultimately galvanized international collaboration to enhance cybersecurity resilience. Below, we examine both the internal response of the affected nation and the broader international community's reaction to this pivotal event.

1) South Korea's Internal Response

The DarkSeoul cyberattack underscored the challenges faced by governments in responding to cyberattacks when the perpetrator is difficult to identify. The repeated cyber assaults on South Korea highlighted the urgent need for both offensive and defensive cybersecurity capabilities. South Korea had already started enhancing its cybersecurity posture before the DarkSeoul attack, focusing on strengthening strategic resources. Recognizing the rising cyber threats, the government began allocating sufficient institutional and financial resources to bolster its defenses. In January 2010,

South Korea established the Cyber Command, and by March 2011, the Defense Ministry had set up a Cyber Policy Team. The Cyber Command played a central role in combating North Korea's online tactics, including counter-propaganda efforts such as posting on North Korean social media platforms and blocking access to North Korean broadcasts (Martin, 2016; Cybersecurity and Infrastructure Security Agency, 2013; Carnegie Mellon University, 2013). In 2013, the Cyber Policy Department was established to monitor cyberspace and telecommunications. That same year, South Korea announced an ambitious plan to train 5,000 cybersecurity experts and doubled its cybersecurity budget, aiming to reach $8.76 billion by 2017 (Martin, 2016; Kshetri, 2014; Kshetri, 2016; Kim & Polito, 2019).

2) International Community's Reaction
The international community responded to the DarkSeoul attack with growing concern over the increasing prevalence of state-sponsored cyber warfare. Although directly attributing the attack to North Korea posed significant challenges, many nations and cybersecurity experts condemned the incident for its devastating impact on critical infrastructure. Organizations like the United Nations pointed to the attack as part of a broader trend of North Korea's cyber aggression. In the aftermath, South Korea intensified its collaboration with international allies, focusing on enhancing its cybersecurity frameworks (National Cybersecurity and Communications Integration Center, 2013; Baezner, 2018). This cooperative response not only strengthened South Korea's defenses but also emphasized to the global community the importance of collective action in countering state-sponsored cyber threats. Collaborative efforts in intelligence sharing and policymaking were critical in shaping South Korea's defense strategies following the DarkSeoul attack. The incident highlighted the necessity for nations to stay vigilant in the face of evolving cyber threats from state actors (CISA, 2013; Carnegie Mellon University, 2013; National Cybersecurity and Communications Integration Center, 2013; Hwang & Choi, 2021).

**Techniques Applied in DarkSeoul**
The DarkSeoul cyberattack, a destructive cyber operation attributed to North Korean threat actors, demonstrated a complex, multi-stage infection scenario. Attackers employed a combination of spear-phishing campaigns and malware, sending emails embedded with malicious Trojan downloaders to various organizations. These emails tricked employees into downloading malware, often disguised as legitimate files like SimDisk.exe. The initial infection stage used a component called "Cast Off," which functioned as a downloader designed to bypass security defenses by masquerading as a harmless application. Upon execution, "Cast Off" established persistence mechanisms by modifying registry entries or adding startup scripts, ensuring its continued presence on infected systems. It then retrieved additional malicious payloads, including wiper malware that targeted both Windows and Unix-based systems. These wipers were designed to overwrite the Master Boot Record (MBR) or critical system files, rendering systems unbootable and causing extensive data loss (Prince, 2013; Martin, 2016; Ji-Young, 2019). In particular, the MBR wiper targeted the first 256 megabytes of the hard drive, which contains essential boot instructions needed to start the operating system. By overwriting this portion of the drive with random data, the malware destroyed the system's ability to boot, making recovery extremely difficult without specialized tools. Beyond simple downloading, "Cast Off" often fetched seemingly benign files, such as JPEG images, that contained hidden malicious code. This code was unpacked and decrypted by "Cast Off" and then executed from the system's temporary directory. A key secondary payload downloaded by "Cast Off" was the "Castdos" Trojan, which was used for Distributed Denial of Service (DDoS) attacks on targeted networks (Martin, 2016; see also Marpaung & Lee, 2013; Pasqualicchio, 2019; Klingner, 2021).

## A. Spear-Phishing and Malware Delivery

In addition to spear-phishing, DNS poisoning played a crucial role in redirecting traffic from legitimate South Korean banking websites to fake servers controlled by the attackers, allowing them to harvest sensitive user data and disrupt financial services. The attack also incorporated advanced tactics, such as compromising patch management systems (e.g., AhnLab) by using stolen credentials to distribute malware across entire networks. These techniques allowed the attackers to deliver destructive payloads en masse. Deceptive tactics, including the use of fake hacktivist groups like the "Whois Crew," misled investigators and created false narratives, adding an extra layer of complexity to the attack. Through iterative processes, "Cast Off" continuously downloaded, unpacked, and executed new malicious components, expanding the attack's reach and impact (Prince, 2013; Martin, 2016; Ministry of National Defense, Republic of Korea, 1972).

## B. Advanced Tactics: Exploiting Vulnerabilities

The scale and impact of the attack, which compromised both Windows and Unix systems, illustrated its broad scope, affecting multiple layers of infrastructure. Despite its relatively lower technical sophistication compared to advanced malware such as Stuxnet, the combination of spear-phishing, DNS poisoning, and data-wiping malware made the DarkSeoul attack devastatingly effective. It underscored the urgent need for robust cyber hygiene practices, especially improved patch management protocols, to prevent future breaches of this magnitude (Prince, 2013; Martin, 2016; Sanger et al. 2017).

## C. Possible Attack Vectors

Several theories exist regarding how the Lazarus Group managed to execute the attack. One theory, published by Avast, suggests that the main attack vector was the Korea Software Property Rights Council (SPC) website. According to this theory, DarkSeoul was a watering hole attack, a cyber-attack that infects frequently visited websites to compromise user machines. The attackers compromised the SPC website and used it to redirect the victims' computers to another site, controlled by them, using a malicious JavaScript script. The attackers then exploited a known vulnerability in Internet Explorer, **CVE-2012-1889**. This vulnerability in Microsoft XML Core Services 3.0, 4.0, 5.0, and 6.0 allowed the attackers to run code remotely by accessing unused memory locations on the targeted machines (BetaFred., 2022; CISA, 2013; Boo, 2018). There is also a theory that the attackers took control of a patch management server used by many future victims to download updates. This theory asserts that the primary attack vector was a server running patch management software made by AhnLab. The attackers likely gained access to this server either by obtaining credentials through spear-phishing or by compromising other systems and discovering stored data. Another theory, proposed by F-Secure, a Finnish cybersecurity firm, states that spear-phishing and email were the primary infection vectors. The attackers used long file names and double file extensions to trick users into opening infected files. F-Secure claims that the malware disguised itself as a fake Internet Explorer lookalike, tricking users into running it. The malware then targeted the System32 directory on Windows machines and executed its malicious code using a DLL file. Given the size of the attack and the numerous theories proposed by various organizations, it is reasonable to conclude that DarkSeoul was most likely a multi-vector attack, utilizing a variety of technologies. This is further supported by the involvement of a hacker group associated with the North Korean government. As a nation-state, North Korea has the resources and manpower to carry out such a complex attack, capable of targeting various computer systems and infrastructure.

| Metric | DarkSeoul (2013) | WannaCry (2017) | NotPetya (2017) | Stuxnet (2010) | Shamoon (2012) |
|---|---|---|---|---|---|
| Sophistication | Moderate (known malware reuse) | High (used EternalBlue exploit) | High (EternalBlue & Mimikatz) | Extremely high (4 zero-days) | Moderate (destructive malware) |
| Scale (systems affected) | 48,000 systems | 300,000 globally | 300,000 globally | 1,000 nuclear centrifuges | 35,000 computers |
| Impact | Disrupted South Korean banks, media | Global service disruption | Global economic damage ($10B) | Physical destruction | Destroyed data at Saudi Aramco |
| Economic Loss | $800 million | $4 billion | $10 billion | Undisclosed (nuclear damage) | Economic losses for Saudi Aramco |
| Propagation Method | Spear-phishing, DNS poisoning | Worm, ransomware, Eternal Blue | Ransomware, worm | Worm, USB drives | Malware |
| Unique Aspects | Reused old malware, region-specific | Fast global spread, ransom demand | Wiper disguised as ransomware | Targeted industrial control systems | Targeted oi industry |

Table 1: Comparison of DarkSeoul with Major Cyberattacks

D. Tactics Used in the Attack

The attackers employed several tactics to spread their malware and execute the attack. Social engineering was a key element of their operation. Spear-phishing was used to steal important credentials and gain access to critical systems like servers. Another tactic involved exploiting software bugs to access systems and execute malicious code. Once the attackers obtained the necessary information, they destroyed the infected systems' data to cover their tracks, making it difficult for investigators to determine exactly what happened and who was responsible. DNS poisoning, which redirects users to malicious websites, was another important tactic. The attackers likely used DNS poisoning to steal information or exploit the **CVE-2012-1889** vulnerability to run malicious code on the victims' computers (BetaFred., 2022; CISA, 2013; Novetta, 2016).

E. Technologies Used in the Attack

The DarkSeoul cyberattack relied on three main pieces of malware to achieve its goals. The first was a dropper Trojan, designed to download additional malware onto the victim's computer. The second was a Master Boot Record (MBR) wiper, a destructive payload downloaded by the dropper Trojan that erased the data on infected PCs, rendering them unusable. The third was a Remote Access Trojan (RAT), used to compromise a patch management server. The RAT allowed attackers to control the patch management server and distribute the dropper Trojan and MBR wiper to other systems.

F. Dropper Trojan

Multiple versions of the dropper Trojan were employed in the DarkSeoul attack, targeting both Windows and Unix-based systems. The Windows version included executable files that placed malicious payloads in the %Temp% directory, bypassing detection. It also sought out antivirus programs commonly used in South Korea, such as AhnLab and Hauri, attempting to disable them by terminating related processes. Although initially tailored for South Korean antivirus software, experts suggest the malware could easily be adapted to target any antivirus product worldwide. Each variant of the dropper Trojan had unique payloads. For example, Dropper A contained both Windows and Unix components. On Unix systems, it used a Bash script to wipe hard drives by targeting the initial sectors of each partition, including the MBR, rendering the system unbootable. In Windows, Dropper A deployed a wiping executable designed to erase the hard drive and disable the system after shutting down any running antivirus processes. Dropper B included a similar Windows wiping executable but focused on disabling different antivirus programs before proceeding with the wipe. Dropper C operated by injecting its code into system memory, bypassing traditional antivirus detection.

### G. MBR Wiper

The MBR wiper targeted both Windows and Unix-based systems like Linux and Solaris. On Windows, it wiped the MBR partition on each connected hard drive, then shut down the system, preventing it from booting. The Unix version used the "dd" shell command to wipe system data. Normally used for routine tasks like system backups, the "dd" command can cause irreparable damage if misused. In this case, the command was deliberately used to delete critical directories on the root partition of a Linux system, making it non-functional.

### H. Remote Access Trojan (RAT)

To distribute the dropper Trojan and MBR wiper, the attackers used a Remote Access Trojan (RAT). This type of malware acts as a backdoor for attackers, giving them control over infected computers. In this case, the RAT was deployed on a server running patch management software developed by AhnLab. It is suspected that the attackers used this server to distribute their malware to other systems using the same patch management network.

### I. Attack Coordination and Procedures

The DarkSeoul attack was highly coordinated, demonstrating meticulous planning and precision. The attackers effectively conducted a spear-phishing campaign to access numerous vital systems, collecting a vast amount of sensitive data from the South Korean government and major financial institutions. The destruction of these systems played a critical role in the overall attack, allowing the attackers to cover their tracks and make it harder for investigators to trace the attack back to them. The attack appeared to serve dual purposes: gathering intelligence and concealing previous cyber operations against South Korea. While the attackers initially focused on data collection, the destructive nature of the attack suggests an underlying motive to cripple critical infrastructure by wiping systems and rendering them non-functional.

### Comparison with Other Major Cyber Attacks

While the DarkSeoul cyberattack was highly disruptive within its targeted region, it is essential to place its impact, scale, and sophistication within the broader context of other significant nation-sponsored cyberattacks. By comparing it to other well-known attacks like WannaCry, Not-Petya, Stuxnet, and Shamoon, we can better understand

DarkSeoul's role in the evolution of cyber warfare and its standing among other landmark threats. Each of these attacks offers different insights into the strategies and objectives of state-sponsored hackers, ranging from espionage to outright destruction of critical infrastructure.

## A. WannaCry (2017)

WannaCry was a global ransomware attack that affected over 300,000 systems across numerous regions. It spread rapidly using the EternalBlue exploit, a vulnerability in widely-used operating systems. The WannaCry attack primarily aimed to encrypt users' data and demand a ransom in cryptocurrency for its release. Although the attack did not specifically target critical infrastructure, its widespread reach caused massive disruptions in hospitals, businesses, and public services globally. Critical healthcare services in particular were significantly impacted, forcing cancellations of surgeries and diverting emergency patients.

In terms of scale, WannaCry dwarfs DarkSeoul in the number of systems affected and the geographical reach of the attack. However, while WannaCry caused chaos globally, it was less destructive than DarkSeoul because it did not aim to destroy data or disable critical systems permanently. DarkSeoul's focus on wiping data from critical financial institutions and media companies, combined with its targeted attack on the region's infrastructure, makes it more strategically focused and devastating in its local impact compared to the global, non-discriminatory spread of WannaCry.

## B. NotPetya (2017)

Like WannaCry, NotPetya was another devastating attack that occurred in 2017. While it initially masqueraded as ransomware, its true objective was destructive. NotPetya targeted corporations during a period of heightened political tension in a specific region, but the malware quickly spread beyond its initial target, infecting global corporations. NotPetya exploited the same EternalBlue vulnerability as WannaCry but was far more destructive—rendering systems inoperable by overwriting the Master Boot Record (MBR), which controls system startup. Companies in various industries were hit, leading to $10 billion in global economic damages. While NotPetya caused substantial global damage, its impact on its initial target was similar to the DarkSeoul attack in its focus on destabilizing critical infrastructure. However, the global spread of NotPetya to major international companies magnified its overall impact far beyond its original regional borders. In contrast, DarkSeoul's primary impact remained confined to its targeted region, although the attack on media and financial sectors had profound consequences for internal stability and public trust in infrastructure.

## C. Stuxnet (2010)

Stuxnet is often cited as the first known instance of a cyberattack that caused physical destruction. Widely believed to have been developed jointly by two collaborating nations, Stuxnet specifically targeted nuclear infrastructure in a particular region. The malware spread through infected USB drives and was designed to damage centrifuges used in uranium enrichment by causing them to spin out of control while providing normal readings to operators. Stuxnet's objective was clear: sabotage nuclear ambitions in its target region. In terms of sophistication, Stuxnet is unparalleled. It exploited four zero-day vulnerabilities in widely-used operating systems, a rarity in cyberattacks. Stuxnet was custom engineered to target industrial control systems (ICS), which made it highly specialized. While DarkSeoul employed known malware and more conventional attack vectors like spear-phishing and DNS poisoning, Stuxnet introduced a new frontier of cyber warfare where digital tools were used to cause real-world, physical damage. While Stuxnet had a clear military objective, DarkSeoul was designed more for psychological and economic disruption, targeting banking and media systems to sow panic and destabilize public trust. The two attacks differ in their focus—Stuxnet was a strategic military strike, while DarkSeoul was a nationwide disruption of civilian infrastructure—but both highlight the growing role of cyberattacks in achieving state-sponsored goals.

D. Shamoon (2012)

Shamoon, like DarkSeoul, was a destructive cyberattack designed to wipe data. Shamoon targeted the energy sector and infected 35,000 computers, effectively crippling operations for days. The malware wiped files and replaced them with politically charged images. Shamoon was likely motivated by regional political tensions and aimed at destabilizing the economy by disrupting critical energy production capabilities. When comparing Shamoon to DarkSeoul, both attacks share similarities in their destructive intent. Both targeted specific sectors of a country's economy and sought to destroy critical infrastructure. However, while Shamoon aimed at a single entity (in this case, an oil company), DarkSeoul targeted a broader spectrum of financial and media institutions, making it more widespread in its impact on civilian life. In terms of economic loss, Shamoon's attack had a significant impact on the global energy market, while DarkSeoul's focus was on the domestic economy and public infrastructure.

E. Comparison Results

In comparing DarkSeoul to other major cyberattacks, it becomes clear that while it was more regionally confined, it was highly destructive and strategically focused on key infrastructure sectors. Its targeted nature, particularly towards media and financial institutions, makes it distinct from more globally-oriented attacks like WannaCry and NotPetya. Furthermore, DarkSeoul's psychological and economic disruption tactics align it more closely with attacks like Shamoon, though it lacked the physical destruction characteristic of Stuxnet. DarkSeoul's significant impact on critical infrastructure and the public panic it generated underscore the evolving tactics in nation-sponsored cyber warfare. Though it didn't reach the global scale of attacks like NotPetya or WannaCry, its focused disruption of an entire country's essential services highlights the potential for future regional cyber conflicts to cause widespread damage without necessarily spilling over to other countries. The attack also emphasized the need for nations to strengthen their cyber defenses and to cooperate internationally in the face of increasingly sophisticated and destructive state-sponsored cyberattacks.

## Our defense recommendations

Considering the pervasive use of phishing in the DarkSeoul attack, it is evident that a substantial portion of the attack was predicated on extracting sensitive information from unsuspecting users. The DarkSeoul incident is not an isolated case; numerous other cyber threats heavily rely on similar tactics. In fact, it can be reasonably asserted that a vast majority of cyber threats involve phishing and social engineering in some capacity. This leads to the conclusion that the most critical vulnerability in this cyberattack was the human element—employees working within the targeted organizations. Although the attack primarily exploited software vulnerabilities, its success hinged on the inability of key personnel in critical institutions to recognize that they were being deceived into divulging confidential information to unauthorized parties or opening email attachments that were malicious in nature. We contend that the success of this attack can largely be attributed to insufficient cybersecurity training within the affected organizations. It is imperative that all personnel handling sensitive computer systems and information receive comprehensive training in identifying common cyber threats. We recommend that organizations dealing with sensitive data implement thorough training programs, educating their employees on how to detect deceptive emails by providing instruction on identifying phishing attempts, and illustrating the potential consequences of failing to do so. Additionally, organizations should enforce a stringent email communication policy, mandating that all legitimate com- pany emails follow a standardized format and are used exclusively for work-related communications. Moreover, employees should be discouraged from sending or opening files attached to emails, as more secure methods of file sharing, such as FTP servers, should be employed within the organization. Organizations must also perform regular security audits in collaboration with reputable cybersecurity consulting firms. These audits should be conducted covertly to properly assess the adherence of employees to established cybersecurity protocols, thus enabling organizations to accurately identify their vulnerabilities. Furthermore, it is essential that organizations employ an internal cybersecurity specialist with substantial expertise in identifying and responding to cyber threats in real time. This is crucial, as the delayed detection of the DarkSeoul attack allowed the Lazarus Group to compromise thousands of systems over an extended period. While this is not an exhaustive set of recommendations, it represents a pragmatic starting point that can be swiftly and efficiently implemented within any organization. The mitigation of software exploits used in such attacks presents a more complex challenge. No software is entirely immune to vulnerabilities, and most professional software is highly intricate, making it difficult to address flaws comprehensively. In many cases, resolving a particular vulnerability requires significant portions of the software to be rewritten from scratch. Therefore, to prevent attackers from exploiting software weaknesses, developers must adopt a proactive and security-centric approach throughout the software development lifecycle. In addition, Software developers must rigorously evaluate their code to ensure that it cannot be exploited for unintended purposes. Continuous and systematic checks throughout the development process are necessary to guarantee that the software is being developed securely from inception to completion. In addition, there must be more stringent regulatory oversight regarding software security. Large software companies, such as Microsoft, possess the requisite resources to prioritize the remediation of critical security flaws, yet often fail to allocate sufficient personnel to address these issues. This is largely due to the absence of legal mandates requiring companies to fix security vulnerabilities, which diminishes the perceived urgency of such issues. The obvious remedy for this problem would be to transition to a software provider that places a greater emphasis on security. However, in many cases, this is impractical due to the lack of viable alternatives for certain professional software. This dynamic grants significant leverage to software developers over both their users and the market, enabling them to neglect the remediation of vulnerabilities. While Microsoft is not the sole offender, it is certainly one of the most prominent. Given the lack of competition among many software companies, the only viable mechanism to compel them to address critical vulnerabilities is through the enactment of legislation mandating the prioritization of security patches. Furthermore, we propose that software service providers, such as Microsoft, offer real-time cybersecurity defense services to users who have purchased licenses for their products during critical cyber threats. For instance, in the DarkSeoul attack, the malware attempted to execute critical operations such as terminating processes and receiving remote code execution commands, as discussed in this paper. We recommend that Microsoft provide 24/7 online monitoring services for all licensed software connected to the network. Such a system would ensure that if a device exhibits signs of compromise, an immediate alert is sent to the support team through Microsoft's Windows Security plugin, allowing organizations to be notified of the attack in real time rather than retrospectively. The implementation of such mechanisms would facilitate timely responses, granting organizations critical time to deploy defensive strategies and safeguard compromised systems. By enabling real-time awareness of an attack, organizations can significantly enhance their capacity to mitigate the damage, which is crucial in minimizing the impact of cyberattacks.

**Conclusion**

DarkSeoul cyber attack serves as a critical case study in state-sponsored cyber warfare, showcasing both the impact and complexity of such operations. Orchestrated by North Korean hackers, the attack used techniques like spear phishing, DNS poisoning, and MBR-wiping malware, crippling South Korea's financial and media sectors by disrupting over 48,000 systems. Although less sophisticated than attacks like Stuxnet, the operation highlighted the devastating consequences of exploiting vulnerabilities in critical infrastructure. Attribution was challenging due to anonymized infrastructures, but the Lazarus Group was identified as the likely perpetrator. The attack underscores the need for enhanced cybersecurity measures, including improved patch management, social engineering awareness, and stronger national cybersecurity frameworks. South Korea's response, through the creation of Cyber Command and increased resource allocation, demonstrates the importance of proactive defense strategies. The global response, led by international bodies like the United Nations, emphasized the importance of intelligence sharing and international cooperation to combat state-backed cyber threats. Ultimately, the DarkSeoul at- tack serves as a reminder of the need for continuous vigilance and collaboration against evolving cyber threats. The lessons learned should guide future efforts to strengthen resilience, particularly in protecting critical infrastructure.